\documentclass[screen,manuscript]{acmart}
\usepackage{array}
\AtBeginDocument{%
  }

\copyrightyear{2026}
\acmYear{2026}
\setcopyright{cc}
\setcctype{by}
\acmConference[ASSETS '26]{The 28th International ACM SIGACCESS Conference on Computers and Accessibility}{October 25--28, 2026}{Vila Nova de Gaia, Portugal}
\acmBooktitle{The 28th International ACM SIGACCESS Conference on Computers and Accessibility (ASSETS '26), October 25--28, 2026, Vila Nova de Gaia, Portugal}
\acmDOI{10.1145/3797867.3829005}
\acmISBN{979-8-4007-2521-0/2026/10}

\begin{document}

\title[Seeing the Voice, Preserving the Self]{Seeing the Voice, Preserving the Self: A Participatory Design Approach to Deaf-Centric Text-to-Speech}


\author{Shela Atemnkeng}
\email{ela.atemnkeng@gallaudet.edu}
\orcid{0009-0002-9241-8537}
\affiliation{%
  \institution{Gallaudet University}
  \city{Washington}
  \state{D.C.}
  \country{USA}
}

\author{Patrick Boudreault}
\email{patrick.boudreault@gallaudet.edu}
\orcid{0000-0002-1101-311X}
\affiliation{%
  \institution{Gallaudet University}
  \city{Washington}
  \state{D.C.}
  \country{USA}
}

\author{Paige DeVries}
\email{paige.devries@gallaudet.edu}
\orcid{0009-0009-9815-7121}
\affiliation{%
  \institution{Gallaudet University}
  \city{Washington}
  \state{D.C.}
  \country{USA}
}

\author{Lloyd May}
\email{lloyd@ccrma.stanford.edu}
\orcid{0000-0003-4692-8261}
\affiliation{%
  \institution{Stanford University}
  \streetaddress{660 Lomita Ct}
  \city{Stanford}
  \state{California}
  \country{USA}
  \postcode{94305}
}

\author{Christian Vogler}
\email{christian.vogler@gallaudet.edu}
\orcid{0000-0003-2590-6880}
\affiliation{%
  \institution{Gallaudet University}
  \city{Washington}
  \state{DC}
  \country{USA}
}
\renewcommand{\shortauthors}{Atemnkeng et al.}

\begin{abstract}
We describe a participatory design approach toward developing Deaf-centric text-to-speech (TTS) technologies. While TTS is growing rapidly in the mainstream, it has received little attention to date in the deaf and hard of hearing (DHH) technology space. Critical problems have remained unaddressed for DHH users, including the ability to manipulate tone, emotions and delivery via non-auditory means. Verifying that the generated speech matches intent and is appropriate for a given situation without having to listen to it is another challenge. Respecting cultural and identity factors in the generated speech is also important. This work explores the design space with DHH participants through two focus groups, three co-design sessions, and four one-on-one early-stage design evaluation sessions. Participants included people both familiar and unfamiliar with TTS, as well as DHH content creators. We describe key findings, design ideas, results, and implications for future Deaf-centric TTS development. We also identify unmet technology requirements that pose barriers to adoption of Deaf-centric TTS technology.
\end{abstract}




\maketitle

\section{Introduction}
In the area of Deaf-centric technology, sign language technologies (recognition, translation and generation)~\cite{efthimiou2009sign} and automatic speech recognition (ASR)~\cite{arora2012automatic} have both received a lot of attention. Rapid technological and artificial intelligence (AI) advances in these areas have transformed the ways deaf and hard of hearing (DHH) people communicate, create media, and participate in professional and social spaces. In this paper, we explore a neglected area that is undergoing a rapid technological evolution: text to speech (TTS). 

TTS is common in the hearing world for dubbing videos, voice cloning, and more~\cite{perez2021towards}. It also has a place as an accessibility solution in the speech disability space~\cite{weinberg2025robot,king2020growing}. Some DHH people use TTS as a mobile communication tool, including generating speech output in everyday apps and using telecommunications relay services~\cite{prnewswireNagishWhich}.

Currently, there are barriers to full adoption of TTS by DHH people. Existing platforms were not designed with DHH users in mind, and their capabilities remain largely inaccessible for two reasons. First, many DHH users cannot reliably assess speech output due to limited auditory access, making it difficult to determine whether the generated speech sounds appropriate, natural, or emotionally aligned with its intended meaning~\cite{rodolitz2019accessibility}. Second, existing systems rely on generic voices that do not reflect personal identity or vocal characteristics, which can be especially important for DHH users seeking representation that feels authentic while still intelligible to hearing listeners. These challenges highlight the need for systems that support non-auditory verification, assurances of voice intelligibility, emotional control, and identity-centered customization.

As an anecdotal data point, a participant in the research described in this paper was happy to show off their use of TTS technology, only to receive feedback from our team that the output was of poor quality (sounding robotic and missing human-like vocal qualities) and may have been perceived negatively by hearing conversation partners. There is a need to understand how TTS technologies can be adapted to support Deaf-centric use cases and --- critically --- give DHH people tools for quality control.

There are many potential use cases for Deaf-centric TTS. In a hearing-and voice-centric world, the lack of accessible speech generation options creates barriers across multiple domains. DHH content creators report reduced engagement on social media when their videos lack voiceovers~\cite{cao2023sparkling,cao2024voices}. Academic and professional contexts increasingly require prerecorded presentation videos, while DHH presenters cannot easily generate intelligible audio narration alongside signed content~\cite{vogler2025barriers}. In everyday communication, whether in person, online, or over the phone, DHH individuals who choose to communicate with speech may encounter challenges with intonation, stress, rhythm, and pitch~\cite{lee2021speech,kang2010suprasegmental}, all of which convey emotion, personality, and sociolinguistic nuance~\cite{leongomez2021voice}. Challenges with interrupting, interjecting, or holding the conversational floor~\cite{napier2011difficult,vogler2013mixed} are reflected in situations where DHH individuals are cut off by others and where they unintentionally overlap due to limited access to auditory turn-taking cues. These further contribute to communicative inequities~\cite{halbe2012s,romero2021visual}. Alternative tools, such as chat features or hand raising on online platforms, remain slower and less effective~\cite{rui2022online}. Even with interpreters, users face issues such as time lags, processing delays, and footing or positionality concerns where the intended stance or alignment of the DHH speaker is altered or misunderstood due to interpreter mediation~\cite{cokely1992interpretation,cokely2005shifting,han2020empirical}.

In this paper, we describe work-in-progress that attempts to build Deaf-centric TTS technology via participatory design, with the goals of providing full control over delivery, tone, and emotions, and enabling Deaf-centric voice cloning capabilities. We specifically describe the methods and findings from focus groups, co-design activities, and early-stage design evaluation with DHH users, hoping to guide the development of Deaf-centric TTS applications in future work.

This work surfaces design requirements, ethical considerations, and user‑defined criteria for appropriate and authentic voice output in Deaf-centric TTS. These insights outline a design space that can guide future technical development while inviting the broader HCI community to engage with the challenges of creating TTS systems that respect Deaf identities, support non-auditory verification, and expand communicative autonomy. This design space provides a foundation for future research and prototyping of Deaf-centric and other identity-aware technologies.

Voice cloning is a topic that requires careful attention. Some DHH users may wish to use voice cloning to faithfully replicate their own voice, for example, because using speech is exhausting for them. Others may wish to use it to reduce or remove the elements of deaf-accented speech that may introduce challenges to DHH people's speech being understood by some listeners. But these goals must always be user-driven: the purpose of voice cloning is not to ``correct'' deaf speech or force DHH users to conform to hearing-centric expectations. Rather, voice cloning capabilities should expand DHH agency by supporting a range of preferences and communicative goals, without presuming any single desired outcome. This aligns with broader work on accent-modification in TTS, where users with non-standard or multilingual accents may similarly seek tools that support intelligibility without erasing identity~\cite{FELPS2009920}.

\subsection{Positionality Statement}
This paper is rooted in the frustrations that members of the research team have experienced firsthand in working with TTS technology, such as creating voiceovers for content. The first author is a deaf graduate student with a developing research interest in accessible technology and inclusive design, uses spoken and written English as their primary modes of communication, and uses caption technologies for communication. The second author is a Deaf professor in sign language studies who knows several sign languages. They rely on AI technologies for everyday accessibility needs, including TTS. The third author is a hearing graduate student with experience in accessible HCI research and a graduate degree in ASL/English Interpreting. They use ASL as well as spoken and written English as primary communication methods. The fourth author is a hearing researcher with experience in accessible HCI research and a PhD in Music Technology. They use spoken and written English as primary communication methods and are conversational in ASL and South African Sign Language. The last author is a deaf professor who uses ASL, as well as spoken and written English and German as their primary modes of communication, and frequently runs into situations where their deaf accent forces them to use TTS or sign language interpreters for voicing.

\section{Related Work}
AI-driven text-to-speech technologies have advanced rapidly. Tools such as OpenAI’s Voice Engine offer broad applications, such as representing text as auditory speech using AI~\cite{openaiNavigatingChallenges}. However, these tools are  designed primarily for hearing users, with little consideration of how DHH users might evaluate or control the output without auditory access. Commercial platforms, including ElevenLabs.io, Lovo.ai, and Speechify.com, claim to offer natural-sounding synthetic speech and customizable voice styles~\cite{elevenlabsFreeVoice,lovoLOVOFree,speechifySpeechifyFree}. Although these platforms provide a range of voice options, their customization interfaces assume that users can aurally verify output quality, a barrier that renders many of their features effectively inaccessible to DHH users. These technologies show promise for supporting multimodal communication and emotional expressiveness~\cite{gessinger2022cross,hillaire2019humanising} and can provide broader accessibility benefits, including improved conveyance of emotional undertones in text-based communication~\cite{schroeter2002perspective,seita2021deaf}. However this promise has not been extended to DHH populations in any systematic way.

Contemporary TTS systems rely on generative AI architectures~\cite{huang2022prodiff,wang2023neural}. While recent systems achieve high naturalness \cite{huang2022prodiff}, challenges remain in producing context-sensitive prosody, expressive variation, and conversationally appropriate speech; speech whose timing, intonation, and interactional cues (such as pausing, emphasis, and signaling when a speaker intends to pause or yield the floor) adapt dynamically to discourse context and turn-taking demands. Emotional TTS research focuses on generating speech that is aligned with predefined categories, such as happiness, sadness, anger, or excitement. Models have achieved progress in capturing broad affective states, but remain limited by simplified emotion taxonomies, limited coverage of mixed or subtle effects, and inconsistent perceptual accuracy across listeners and datasets~\cite{aoki2022clear,shaikh2010improving,zhou2022speech,gessinger2022cross}. Critically, these systems depend entirely on auditory evaluation to assess emotional fidelity, an approach that is inaccessible to many DHH users and that our work directly seeks to address through non-auditory verification mechanisms. Speaker-adaptive TTS approaches have demonstrated the ability to retain aspects of a speaker’s vocal identity~\cite{jia2018transfer,biadsy2019parrotron}. In particular, Biadsy et al. \cite{biadsy2019parrotron} explored speech-to-speech conversion for deaf-accented speech, demonstrating the technical feasibility of adapting voice output for non-normative speakers, an important step toward inclusive voice technology. However, the work focused mainly on intelligibility as defined by hearing listeners, reflecting a different set of priorities than those that surfaced in our study. Our work builds on this foundation by treating DHH users as primary stakeholders in defining what constitutes an appropriate and authentic voice output. 

Existing accommodations, including sign language interpreters, highlight why representation and identity alignment matter for DHH users in voice‑based technologies. Interpreters play a crucial role in enabling access, yet their mediation can introduce time lags, processing delays, and footing or positionality shifts that alter how a DHH speaker’s stance or alignment is perceived. These challenges have prompted many DHH individuals to request interpreters whose signing style or cultural background better aligns with their own \cite{deafservicesunlimitedDebunkingMyths}. ``Deaf consumers may request interpreters whose signing style or cultural background is better aligned with their own'' \cite{deafservicesunlimitedDebunkingMyths}. These interpreter‑related concerns mirror the challenges faced in Deaf‑centric TTS, where mismatched or generic voices can similarly distort how DHH users are represented to hearing listeners. As one source notes, ``If an ASL interpreter cannot represent us to the fullest extent possible, maybe they cannot be the ASL interpreter we need them to be''~\cite{substackBeingTransgender}. This connection underscores the importance of voice output that reflects identity, preserves communicative intent, and supports user‑driven control over representation.

Foundational work on DHH speech production further highlights why current TTS systems fall short for this population. Early analyses describe atypical (relative to hearing/non-DHH speech production) prosodic patterns, restricted phonetic inventories, and disrupted timing and stress, all linked to limited auditory feedback~\cite{osberger1982speech}. Subsequent studies show that intelligibility varies depending on listener familiarity and sentence context, underscoring the need for systems that can enhance clarity without erasing identity~\cite{mcgarr1983intelligibility}. Physiological research also documents differences in respiratory coordination and rhythm among deaf speakers, revealing biomechanical factors that shape prosody and phrasing~\cite{hudgins1934comparative}. Taken together, these findings establish that DHH speech characteristics are not deficits to be corrected, but complex identity-linked patterns that any DHH-centered TTS system must be designed to accommodate and preserve, rather than normalize away. These findings reinforce the need for TTS systems that can model atypical prosody while supporting user-controlled adjustments.

Parallel work on automatic speech recognition (ASR) demonstrates the limitations of existing speech technologies for DHH users. Studies consistently show that commercial ASR systems perform poorly on deaf-accented speech, with word error rates exceeding 70\% even for highly intelligible speakers~\cite{glasser2017feasibility}. User evaluations in school and workplace settings reveal that ASR accuracy is highly sensitive to prosodic variability and background noise, limiting its reliability for real-time communication~\cite{gottermeier2016user}. Research on voice-controlled devices similarly finds that ASR pipelines often do not recognize DHH speech,, despite  recent improvements~\cite{devries2026ipa}, prompting users to rely on TTS output as a more reliable alternative~\cite{bigham2017deaf}. These findings show that when ASR fails to recognize their speech, DHH users turn to TTS systems to speak on their behalf, but these TTS systems similarly lack the mechanisms to faithfully represent their identity, tone, and expressive intent, leaving a gap on both ends of the voice communication pipeline.

Previous work in human-computer interaction emphasizes participatory and co-design approaches as essential for technologies for DHH communities. Co-design directly involves prospective users in the development of technologies, where the users' perspective is paired with that of experts~\cite{sanders2008co}, aligning closely with inclusive design practices that aim to create systems that are usable by the widest range of people, including those with non-normative communication patterns~\cite{treviranus2018one,treviranus2018two,treviranus2018three}. These approaches work well together because inclusive design provides the values and goals; equity, flexibility, and variability. Co‑design provides the concrete practices for operationalizing those values with the communities most affected. Sanders and Stappers \cite{sanders2008co} distinguish co-design from consultation-based approaches by emphasizing genuine creative collaboration between users and designers as a means of surfacing richer, more grounded design insights than traditional user studies alone can provide. Co-design has been widely applied in assistive technologies~\cite{weinberg2025one}, including those targeted at DHH users. Past work on co-design has elicited practices for working with a DHH population both in-person~\cite{sales2020participatory,may2024co} and remotely~\cite{seita2022remotely,mcdonnell2023easier}. Recent findings in sign language research further highlights the risk of oversimplifying Deaf communicative practices. Tang and Piper's study of Chinese deaf content creators demonstrates the multicultural and multilingual complexity of Deaf-led translation and shows why automated systems often fail to capture the richness of signed communication \cite{tang2026reimagining}. These findings underscore the importance of designing technologies that respect the communicative nuance of deaf people, rather than flattening it to overly simplified models. However, none of this prior co-design work has addressed TTS as a design domain, a gap that our study directly fills.

\section{Methods}
The study aims to enhance TTS systems to help DHH individuals convey a broader range of emotions, improving their ability to express complex feelings through speech. It seeks to give DHH users independent control over the emotional tone of their speech, reducing reliance on auditory feedback. The study also focuses on allowing users to personalize their voice, reflecting their unique vocal characteristics for more authentic communication. Additionally, it aims to reduce miscommunication by incorporating better emotional feedback mechanisms and promoting inclusive communication by integrating these emotional and personalization features into TTS systems.

The research was conducted in multiple phases to capture different forms of participant input and progressively refine design ideas. The first phase consisted of two remote focus groups, which explored user experiences and emotional needs related to TTS. This phase was intended to surface participants’ perspectives on emotional expression, identity, and communication challenges, forming the foundation for future design work. The second phase involved three co-design sessions (two on-site and one remote), where participants collaboratively translated insights into interface concepts through sketching and discussion. This phase allowed participants to directly shape the design directions and ensure that the emerging concepts reflected lived experiences and preferences. The third phase comprised four one‑on‑one early‑stage design evaluation sessions, in which participants reviewed wireframes created by the research team from the ideas generated in the focus groups and co-design sessions. This phase focused on assessing how the emerging designs aligned with user expectations and gathering targeted feedback to refine clarity, expressiveness, confidence, and identity alignment. Attention was focused on how the UI/UX features might communicate prosody, identity cues, emotional tone, and contextual appropriateness without relying on auditory perception.

Focus group communication was facilitated in ASL by three DHH and one hearing researcher fluent in ASL, with certified ASL–English interpreting provided for non-signing participants. The interpretation was also captured in transcripts via CART captioning. These transcripts served as a primary data source for data analysis, after members of the research team who were fluent in ASL, written English and also spoken English reviewed the full transcripts against audio and video recordings to verify accuracy due to the potential for interpreter error. The co-design and one-on-one evaluation sessions were facilitated entirely in ASL by the researchers, as all participants were fluent in ASL. All parts of the research received ethics approval, and participants were compensated for their time. \textbf{Beyond institutional review board approval, the study was guided by Deaf‑led participatory ethics that emphasized autonomy in communication methods, respect for sign language as an equal and irreplaceable communication modality, and a commitment to recognize and mitigate potential harm throughout the research process.} The team also engaged with a diversity of Deaf perspectives on TTS technology, acknowledging that participants have differing views on when and how speech technologies should be used. These principles shaped decisions around the study design, facilitation, and interpretation.

In total, 27 people participated across all focus groups, co-design sessions and evaluation sessions. Three of the evaluation session participants were previously co-design participants. Table~\ref{tab:participants} summarizes participant characteristics.

\begin{table*}[h]
  \centering
  \caption{Participant characteristics across focus group, co-design and early-stage evaluation studies}
  \label{tab:participants}
  \begin{tabular}{|>{\raggedright\arraybackslash}p{0.14\textwidth}|>{\raggedright\arraybackslash}p{0.14\textwidth}|>{\raggedright\arraybackslash}p{0.14\textwidth}|>{\raggedright\arraybackslash}p{0.14\textwidth}|>{\raggedright\arraybackslash}p{0.14\textwidth}|>{\raggedright\arraybackslash}p{0.14\textwidth}|} \hline
  \textbf{Session Type} & \textbf{Hearing Identity} & \textbf{Gender} & \textbf{Race/Ethnicity} & \textbf{Language Used} & \textbf{TTS Usage/Familiarity} \\ \hline
  \textbf{Focus Group~1 (n=5)} & 2~Deaf; 1~HoH; 1~late-deafened; 1~CI user & 2~male; 3~female; 1~non-binary & 4~White; 1~Native Hawaiian/Pacific Islander & 4~ASL; 1~spoken English & 2~daily; 2~a few times weekly; 1~never \\ \hline
  \textbf{Focus Group~2 (n=6)} & 4 Deaf; 1~HoH; 1~hearing-loss & 2~male; 3~female; 1~genderqueer & 5~White; 1~Black/African American & 2~ASL; 1~PSE, written English, 1~ASL, written English, various other signed and written languages; 1~ASL, spoken and written English & 1~daily; 3~former users; 1~<~{1x}/month; 1~once \\ \hline
  \textbf{Co-Design Session~1 (n=5)} & 5~Deaf & 5~male & 3~White; 1~Black/African American; 1~prefer not to answer & 3~ASL; 1~ASL, other; 1~ ASL, written and spoken English & 5~with experience \\ \hline
  \textbf{Co-Design Session~2 (n=6)} & 6 Deaf & 3~male; 2~female; 1~other & 3~White; 1~Asian; 1~Hispanic; 1~American Indian/Alaska Native & 4~ASL; 2~ ASL and written English & 4~with experience; 2~none \\ \hline
  \textbf{Content Creator Co-Design Session~3 (n=4)} & 4~Deaf & 4~female & 2~White; 1~Hispanic; 1~prefer not to answer & 2~ASL; 1~ ASL, written English; 1~ ASL, written English, PSE & 3~with experience; 1~none \\ \hline
  \textbf{Evaluation Sessions (n=4, 3 returning and 1 new)} & 4~Deaf & 2~male; 2~female & 1~White; 2~Black/African American; 1~Hispanic & 1~ASL; 1~ASL, spoken, written, and signed English; 1~ASL, written English; 1~ASL, other & 3~with experience; 1~none \\ \hline
  \end{tabular}
\end{table*}

\subsection{Focus Groups}
Two focus group sessions were conducted on Zoom to explore the experiences, expectations, and design needs of DHH users related to TTS technologies. We intentionally recruited both TTS ''experts'' (frequent users and individuals with prior exposure or familiarity with TTS) and ''non-experts'' (DHH individuals who rarely use, or are not familiar with TTS), starting with experts because this is a relatively new design space and we expected them to articulate early design considerations more readily. However, the self‑reported experience of the participants proved to be unreliable and it was only once the sessions began that we realized the groups could not be cleanly separated. As a result, both focus groups included a mix of expertise levels. The focus groups covered topics related to emotional voice modulation and text to speech technology, and both groups received the same content.

The focus group sessions had two goals: (1) gain an initial understanding of potential user needs and (2) understand how TTS is currently perceived in the US-based DHH community. The results influenced the subsequent co-design sessions.

\subsection{Focus Group Procedures}
Each focus group lasted 120 minutes and followed a consistent structure. After orienting the participants to Zoom etiquette and reaffirming informed consent, the moderator introduced the session topics, which included using written communication with hearing people, familiarity and experience with TTS technology, understanding and using AI, voice quality and personalization in TTS, customizing TTS voices, and evaluating and trusting TTS technologies. Discussion prompts were presented in written English via Zoom’s screen‑sharing tools and simultaneously in ASL. Participants discussed each prompt in a group format, question by question, and also shared ideas for improving TTS technologies.

\subsubsection{Focus Group Analysis Methods}
We analyzed the verified transcripts via multi-stage thematic analysis~\cite{braun2006using}. Because there were no prior DHH-related TTS design publications to guide us, we used a bottom-up, inductive coding approach, loosely informed by a DHH UI/UX lens. Two members of the research team independently reviewed each transcript and each generated preliminary codebooks that reflected the salient concepts, concerns, and design needs expressed by participants. Each analyst then identified recurring patterns within and across sessions to revise the coding. Following independent coding, the analysts met to compare and reconcile codebooks, as well as interpretation of concepts, and to collapse redundant or irrelevant concepts. This reconciliation process produced a consolidated codebook and concept categories.

The finalized codebook was again cross-compared across the two focus groups to ensure consistency and to extract the themes related to TTS user experience, design needs, and concerns, which also resulted in initial thematic grouping. We performed a frequency analysis to identify dominant topics, secondary patterns, and less frequent but conceptually meaningful themes. These themes form the basis of the findings reported in Section~\ref{sec:fg-results}.

\subsection{Co-Design Sessions}

The co-design phase built on the focus group findings. The co-design sessions aimed to translate the focus group insights into interface concepts, interaction flows, visual schematics, and control mechanisms, with the goal of informing the development of DHH-centered TTS user interfaces.

We followed a participatory design approach, with DHH participants collaborating directly with researchers to propose, critique, and refine potential designs.  We conducted a total of three co-design sessions; the first two were in-person. In the second session, we introduced personas (cf. Section~\ref{co-design session-procedures}) that we developed from patterns observed in the focus groups and early design discussions. These personas were created to represent distinct communication contexts, user goals, and to help participants reason about how different DHH users might interact with TTS systems. Personas were particularly effective in surfacing nuanced perspectives, which revealed the importance of understanding the content creator viewpoint. To this end, we recruited DHH content creators from across the country and conducted a third co‑design session via Zoom. This expansion allowed us to capture both everyday user perspectives and the specialized needs of those producing content for broader audiences.

\subsubsection{Co-Design Session Procedures}
\label{co-design session-procedures}
The co-design sessions took 150 minutes each, in four phases, which consisted of setting the stage, collaborative design ideation in breakout groups, cross-break\-out group review, and an all-hands final discussion. These are described below.

\textbf{Setting the Stage:} The researchers provided an overview of TTS and potential use cases. This overview included a screening of voiced video clips with captions, and an ASL performance of the associated emotions and delivery style --- which turned out to be different from what the participants had imagined the voicing to sound like. This helped our participants understand the importance of controlling and verifying emotions, style, and delivery in TTS applications. Participants were further asked to discuss prior experiences and expectations related to TTS. Additionally, in co-design session 2, we introduced prospective TTS user personas. Those were (1) a Deaf native ASL signer using TTS for media; (2) a Deaf native ASL signer using TTS for communication, (3) a hard of hearing public speaker using TTS for presentations and speech, and (4) a deaf intermediate signer using TTS for communication with customers. Each persona also featured a description of background, pain points, goals, technology use, and example scenarios. All materials were delivered on slides in written English, and explained in ASL.

\textbf{Collaborative Design Ideation in Breakout Groups:} We split participants into groups of 2--3 people each, plus a group consisting of the researchers. Each breakout group was prompted to generate and discuss TTS UI designs through sketches and drawings, based on the information shared during the setting of the stage. Co-Design Session 2 specifically prompted participants to ideate designs that fit the needs of each of the personas. Co-Design Session 3 asked the participants to come up with designs based on their own firsthand perspectives as content creators. Sessions 1 and 2 provided flip charts and whiteboards for participants to sketch out their ideas, while Session 3 used FigJam~\cite{figmaFigJamOnline} to allow participants to collaborate remotely. The group consisting of the researchers used the same process, but also had their members periodically check in on the participants and provide additional prompts and clarifications as needed.

\textbf{Cross-Breakout Group Review:} Participants reviewed the sketches from the respective other breakout groups. They were given the opportunity to add comments and questions to others’ sketches, and star ideas that they found especially promising. The researchers saved pictures of the sketches and comments. In session 2, we additionally asked each group to explain their sketches in video-recorded ASL to provide additional documentation of their intent.  

\textbf{All-Hands Final Discussion:} For the last 30 minutes, the participants and researchers reconvened to discuss commonalities across the designs, areas of alignment, and divergence. This phase also aimed to identify promising design ideas for eventual prototyping.

\subsubsection{Co-Design Analysis Methods}
For the analysis, we examined photographs of whiteboard sketches, FigJam whiteboards, collections of sticky notes documenting participants' comments, CART transcripts of the all-hands discussions, and video recordings. We condensed the video recordings and transcripts into a write-up of key design ideas while identifying recurrent patterns. The visual artifacts were then cross-referenced with key ideas and discussion data to ensure that design concepts were interpreted within the context in which they emerged. We additionally translated these findings into a list of current and future technical prerequisites (see also Section~\ref{sec:tech-reqs}) for deploying DHH-friendly TTS technologies, across the domains of TTS research, AI research, sign language technology research, and UI/UX work. We also conducted a thematic analysis of the transcripts and recordings to identify additional patterns not captured in the visual artifacts.

\subsection{Early-Stage Design Evaluation}

Following the focus groups and co-design sessions, we conducted a total of four one-on-one design evaluation sessions with DHH participants to understand how DHH users interpret early visual representations of prosody, tone, and delivery in TTS systems. 

We applied the 80/20 rule, also known as the Pareto Principle, to prioritize design ideas generated during the co-design sessions. This principle suggests that roughly 80\% of the value or utility of a system is often derived from 20\% of its features~\cite{obendorf2009minimalism}, p. 92. By focusing on the vital few elements that offered the greatest impact on non-auditory communication, we selected three core visualization concepts for further development, described further in Section~\ref{sec:oneonone} following the co-design results in Section~\ref{sec:co-design results}.

\subsubsection{Short Demo Prototypes Preparation}
To prepare for the evaluation, we translated the prioritized design ideas into wireframes using Figma. These were arranged in a storyboard-style format to support comparison across interaction stages. To enhance visual clarity and reduce cognitive load for participants, wireframes were printed on a large sheet of paper. This eliminated the need for scrolling on a computer and helped participants maintain visual continuity while comparing designs.

We converted the wireframes into short demo prototypes using Figma Make and OpenAI Codex. Figma Make enabled us to scaffold basic interaction flows directly from the wireframes we created, while Codex was used iteratively to generate lightweight web-based short prototypes that simulated typing text and previewing synthesized speech. The prototypes were run locally, with the interface rendered in the browser and the speech output generated through a simple Node.js connection to OpenAI’s text-to-speech service. This allowed participants to experience the intended interaction flow without requiring a fully deployed system.

\subsubsection{Design Evaluation Procedure}
Each session lasted approximately 60 minutes and was facilitated in ASL by members of the research team. Participants were introduced to a scenario designed to ground the activity in a realistic communication context. For example, one of the scenarios asked participants to imagine sending a TTS voice message to their class (to introduce their project) where they wanted the message to sound excited and confident. Participants were asked to consider how they would verify whether the generated voice output matched their intended tone, energy, and style. The evaluation proceeded in three phases, moving from quick intuitive reactions towards deeper and more reflective comparisons.

\textbf{Five-Second Test:} To capture rapid, intuitive impressions of each concept, participants viewed each design briefly and were asked what stood out, what they believed the visualization was communicating, and how they expected it to function. This phase provided insight into the immediate interpretability of prosody cues.

\textbf{Short Prototypes Demonstration:} Participants viewed the short demo prototypes illustrating how each design concept might behave or function in a real-world scenario. These demos included animated examples of pitch, pace, emphasis, and emotional variation (see also Section~\ref{sec:oneonone} for specific UI details). We then asked participants which elements helped them understand how the TTS voice might sound, which design they felt performed best, which cues helped them notice changes in pitch, pace, or emphasis, and how clearly each design conveyed differences in tone, energy, and expressiveness.

\textbf{Side-by-Side Comparison of Designs:} Participants completed a side-by-side walkthrough of all three design concepts using the printed storyboard-style wireframes. They were encouraged to compare the designs at their own pace and reflect on which visualization best  let them judge how their TTS voice output may sound. We used a series of semi‑structured questions, which covered the following topics: (1) which design felt most effective or challenging, (2) how clearly each conveyed changes in prosody, (3) what would make these cues easier to understand, (4) whether any concepts could be merged or improved, and (5) what features they would want if anything were possible.

\subsubsection{Design Evaluation Data Analysis}
\label{sec:design-data-analysis}
All evaluation sessions were video-recorded, while participants' responses and facilitators' observations were documented through typed notes. The evaluation data were reviewed and compared across participants using a goal‑driven comparative approach, focusing on how each person engaged with and interpreted the core design concepts. Facilitator notes and participant comments were synthesized to ensure consistency in how relationships among the four predefined design goals: clarity, expressiveness, confidence, and identity alignment were captured.

\section{Results}
\subsection{Focus Group Results}
\label{sec:fg-results}
We identified six overarching themes through our thematic analysis: (1) access and inclusion, (2) vocal characteristics, (3) communication, (4) identity and personalization, (5) barriers to adoption, and (6) key usability requirements. A total of 152 codes were identified across the dataset by our researchers, providing a foundational structure for organizing and interpreting participant perspectives.

\subsubsection{Access and Inclusion}
Highlighting the need for equity, one participant noted that many DHH individuals use ASL or another first language rather than English, stating, \textit{“Many of us forget… if you need to use text to communicate with a deaf person, they might not know what you’re saying” (P1-5).} Inclusion of DeafDisabled people came up, with participants emphasizing the importance of considering those with \textit{“additional disabilities who are maybe DeafBlind” (P2-3)} who are often left out of technological designs.

\subsubsection{Vocal Characteristics}
One participant noted the importance of acoustic expertise: \textit{“[It] would be helpful if we had someone else to help with modifying.  Kind of like ‘audiologist’ or ‘toneologist.’  [The expert] will need a lot of training and experience for optimal use” (P1-6).} One participant pointed out that manner of speech is important for mutual understanding: \textit{“a lot of the problem with text messages is lack of a tone or expression, and sometimes text messages get misinterpreted” (P1-1).} Others pointed out how tone/delivery would need to be identity-matching, for example \textit{“an androgynous voice to match my own identity” (P2-3, a participant who identifies as genderqueer),} as well as be situation-dependent, so that it \textit{“matches the context of what I’m trying to say and the intent” (P2-1).} For example, \textit{“to pick a white voice over to get some sort of advantage” [instead of a voice of another race/ethnicity to avoid biases] (P2-6)} or, more generally, \textit{“the ability to save specific TTS voices for specific situations” (P2-6).}

\subsubsection{Communication}
The specifics of the communication situation is important, and TTS systems \textit{“need to be adaptable” (P2-1) “based on the text of what is being said… that could be chosen from a drop down list […] pre populated in there based on the actual phrase said” (P1-4).} Communication barriers to using TTS surfaced: \textit{“it just depends on the environment… if I’m in a loud environment… they can’t hear” (P2-2).} Others are hesitant to use TTS in important situations: \textit{“there are still some situations where an interpreter is just a better tool for me and how to describe that to hearing people who don't understand what it is like to be a deaf person, and do my job as a deaf person, and to get them to get it and listen to me” (P2-1).}

\subsubsection{Identity and Personalization}
Participants wanted TTS to represent them accurately, noting that they \textit{“would probably base it on my identity” (P2-2), ``I'm a gay person. I use certain slang associated with my identity. I may want to make customizations that would emulate the vibe of who I am as a person. If I was in a serious situation versus a neutral tone situation or slang...'' (P2-2)} or would choose an identity tailored to the situation \textit{“sometimes I would choose a man’s voice just because sadly in our world it’s still true that people listen/respect a man’s voice more” (P2-3).} It is clear from these comments that intersectional identities need increased attention in this project to allow accurate representation; see also Section~\ref{sec:limitations}. When asked if they would want to use their own voice for TTS, several felt negatively due to privacy concerns and would rather have \textit{“privacy protection built in. And I'm not sure if I would trust that even. So I would more likely opt to use an alternative voice rather than my own” (P1-2).} One participant mentioned hesitation using TTS, noting that \textit{“some people may be offended by that… [and] think we’re not speaking to them directly” (P2-6),} and another weighed the pros of typing instead \textit{“it is easier to control my narrative when I use English by typing back and forth” (P2-6).} Verifying TTS voices is complex, as well: \textit{“It is an interesting question, right?  I'm Deaf. [For me, that means]  not knowing if the sound or voice matches the person” (P1-4).}

\subsubsection{Barriers to Adoption}
Accessibility is important for users with TTS systems. Participants wanted to ensure \textit{“[they] can see the captions and the person they are talking to can see what they are saying” (P1-1).} Compared to showing typed text to someone, or using an interpreter, TTS delays are a concern, since they could cause communication partners to lose patience. \textit{“[Y]ou get a lot of hang ups waiting for that system to work, so it’s not as effective” (P1-4).} ASR for verification uses AI, which some viewed positively as it \textit{“help[s] filter voices and improve accuracy” (P1-3)} whereas others focused on the risks. \textit{“The risks are pretty scary” (P1-2)}, another stated \textit{“when [it comes to] one on one communication, [I use] pen and paper, especially in medical settings” (P1-2)} indicating a preference for tried and true methods. Showing a message for others to read may be simpler than TTS, \textit{“so a person could see what I had typed instead of having to listen to it… most people would probably prefer just quickly reading [my message] instead of listening to it” (P1-3).}

\subsubsection{Key Usability Requirements}
Several participants discussed the importance of how TTS sounds: \textit{“people I worked with didn’t like it because they said it sounded like a robot voice” (P2-3)} and \textit{“people kept saying they did not understand what in the world I was saying… I don't know if it is how the words were pronounced. [...] The voice sounded very mechanical perhaps. Anyhow, it caused a lot of issues to the point where I stopped using it” (P2-6).} Some explored features that would make them more likely to use a TTS app, such as tone-presets and, \textit{“having an offline mode […] if you don’t have internet service” (P2-6).} Another participant stressed the importance of trusting that the technology is working \textit{“I like using notes [...] I don't know if the other person can hear and [...] if it is working or when it is running and when it's not. [...]” (P2-4).} There is a need to retain control of the message through the TTS app, \textit{“how you would modulate a voice, for a different experience, I think depends on the design of the UI” (P1-3).}

\subsection{Co-Design Results}
\label{sec:co-design results}
The co-design sessions revealed a progression of design priorities through personalization, customization, verification and workflows. Overall, they highlighted a strong desire for future TTS systems that would allow DHH users to visually evaluate TTS voice outputs for alignment with intent and preservation of deaf voice identities. These design priorities were shaped by participants' own communication backgrounds and modalities, as reflected in the language-use patterns summarized in Table \ref{tab:participants}.

\begin{figure}[ht]
    \centering
    \includegraphics[width=0.4\textwidth,alt={Minimalist TTS interface layout showing two sections: the top for entering personal attributes like age, sex, race, and accent, and the lower for modifying tone and emotion based on context, such as casual or business-like settings.}]{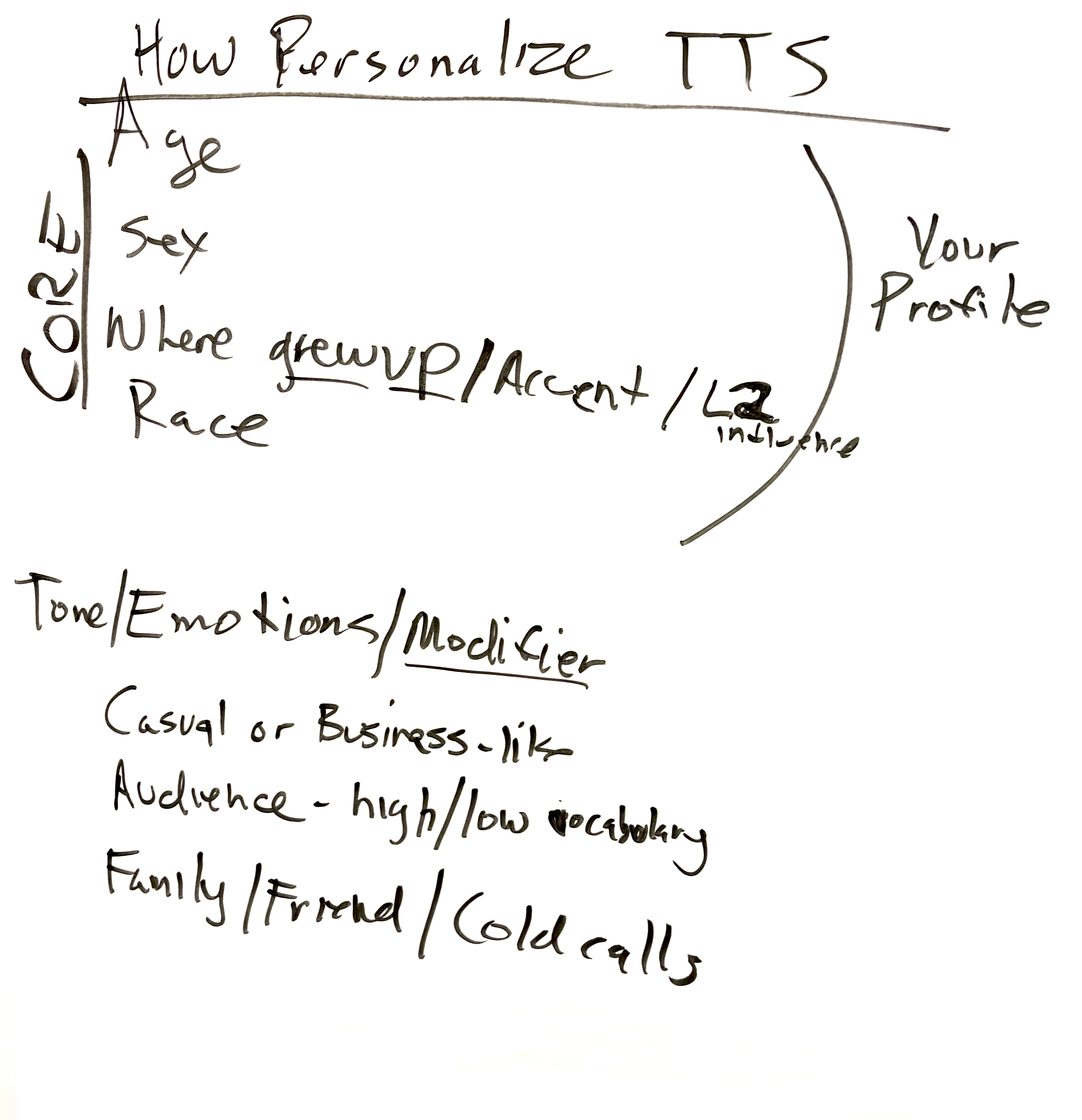}
    \caption{Preset scenarios and simple personalization profiles from participants in the co-design session.}
    \label{fig:presets}
    \Description{Minimalist TTS interface layout showing two sections: the top for entering personal attributes like age, sex, race, and accent, and the lower for modifying tone and emotion based on context, such as casual or business-like settings.}
\end{figure}

\begin{figure}[ht]
    \centering
    \includegraphics[width=0.4\textwidth,alt={TTS output with highlighted words tagged by emojis to convey emotions. A sentence reads “THIS is A TEXT,” the words “THIS” and “A TEXT” are highlighted, and paired with an emoji below to indicate emotional expression, demonstrating how emojis could be linked to specific words to help DHH verify tone visually.}]{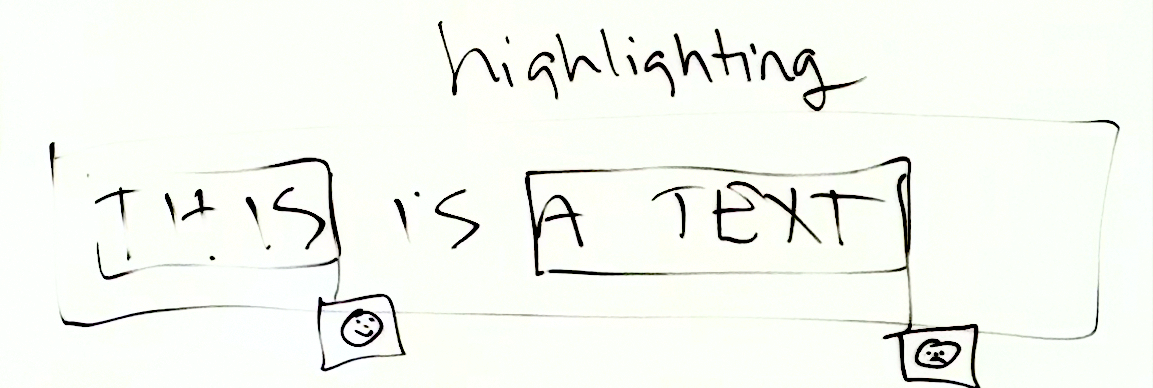}
    \caption{Emojis along a read-out of generated TTS to indicate tone.}
    \label{fig:emojis}
    \Description{TTS output with highlighted words tagged by emojis to convey emotions. A sentence reads “THIS is A TEXT,” the words “THIS” and “A TEXT” are highlighted, and paired with an emoji below to indicate emotional expression, demonstrating how emojis could be linked to specific words to help DHH verify tone visually.}
\end{figure}

In the first session, participants emphasized personal autonomy, as well as personalization through simple presets and features, prioritizing ease of use and quick access (Figure~\ref{fig:presets}). Features with sliders, toggles, and emotional wheels were proposed to allow rapid adjustments of tone without complexity. Preset scenario cards, such as for doctor visits or casual chats, and behavior sliders to shift between tones like serious or ironic, were common across designs. Multimodal verification was suggested through ASL avatars, haptic feedback, audio waveform visualization and emoji cues, with emojis envisioned as small, universally recognizable icons that could appear alongside text to speech output to visually signal the intended tone and emotion; for example, a smiling face could indicate friendliness (Figure~\ref{fig:emojis}). Participants also raised the concept of a ``speaking avatar'', a visual figure that would animate in real time to depict the emotional delivery of the generated speech. These cues were suggested to provide a quick and accessible confirmation of emotional intent without requiring users to listen to the audio. The proposal of ASL avatars, in particular, came predominantly from participants who identified ASL as a primary or preferred communication modality, reflecting a visual-spatial approach to verifying TTS output that paralleled their own communicative practices, rather than relying on text or audio-based confirmation.


In the second co-design session, participants focused on identity‑driven customization and evaluation. Participants proposed tools that emphasized authenticity and trust, including voice profile builders with cultural and demographic markers, expressive avatars and layered emotional tagging. These ideas reflected the understanding that how people speak communicates far more than words alone. Preserving features such as Deaf accents was seen as a critical requirement for ensuring that synthetic voices do not erase individuality.

\begin{figure}[ht]
    \centering
    \includegraphics[width=0.4\textwidth,alt={Phone interface showing lines of typed text on the screen while the front-facing camera detects the user’s facial expressions to assign emotions to the TTS output.}]{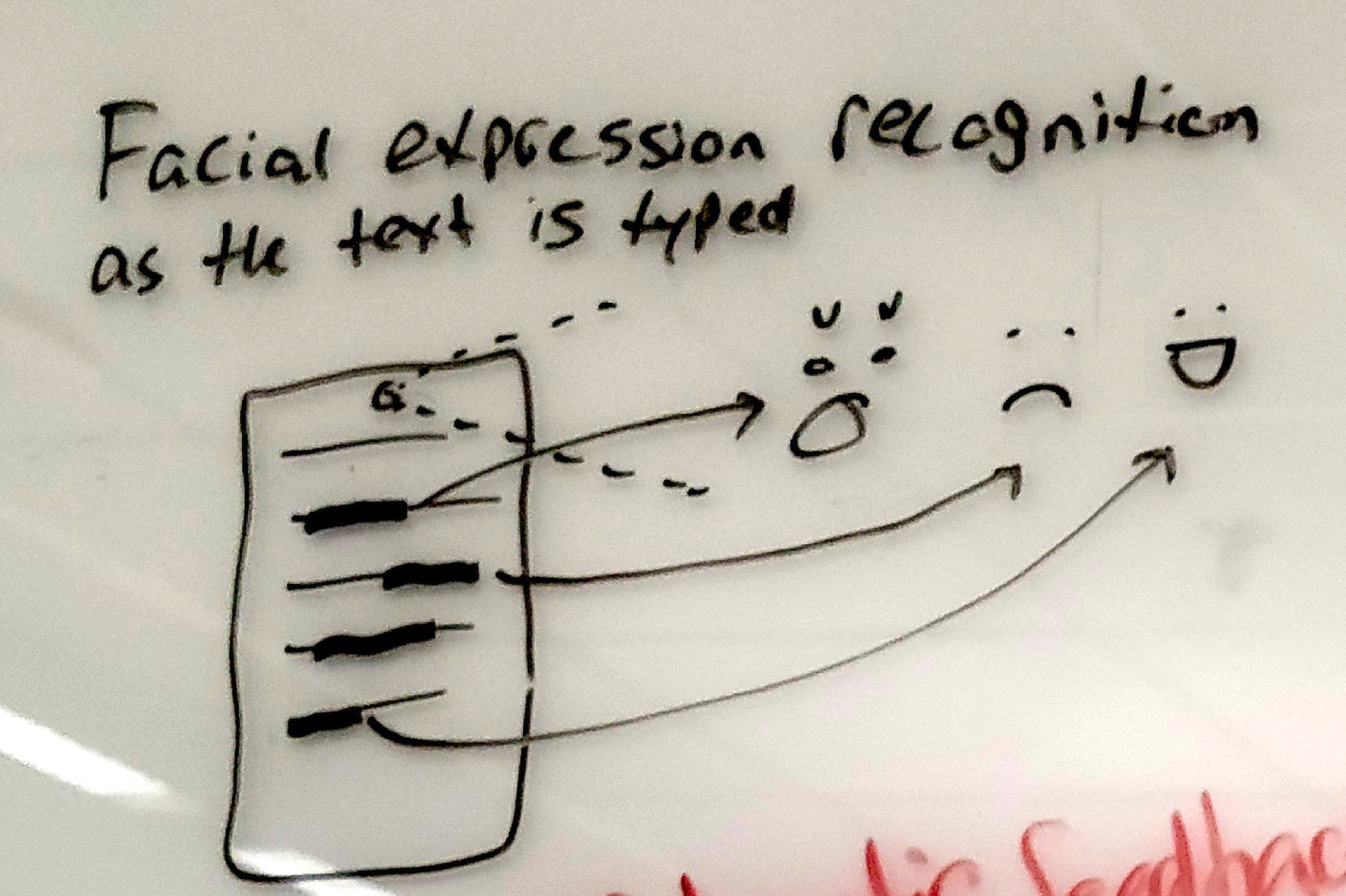}
    \caption{Facial expression recognition of users as they type text, which is to be translated into TTS emotions.}
    \label{fig:facial-expressions}
    \Description{Phone interface showing lines of typed text on the screen while the front-facing camera detects the user’s facial expressions to assign emotions to the TTS output.}
\end{figure}

Paralleling the first co-design session, waveform visuals and scenario‑based modes were introduced to support evaluation, while AI‑generated tone summaries (including an analysis of the emotions, sentiment, and style of speech) offered mechanisms for verifying whether synthetic voices matched user intent. Crowd-sourced audience polling was envisioned where human listeners could respond with thumbs-up or thumbs-down as to whether a generated voice sounded authentic and aligned with the intended tone of the user. Facial expression recognition was suggested to attach tone and emotions directly from the user while typing (Figure~\ref{fig:facial-expressions}). This idea is consistent with the centrality of facial expression as a grammatical component of ASL, where facial markers carry meaning that is inseparable from the signed message itself; for participants who communicate primarily in ASL, facial expression recognition may represent a natural extension of how tone and emotion are already conveyed in their primary language, rather than an unfamiliar new input method. As discussions shifted toward how TTS systems could better represent Deaf identities, rather than just evaluating tone, participants raised the idea of Deaf accent cloning as another way to ensure cultural and identity‑based representation. This emerged from concerns that synthetic voices often default to hearing‑normative speech patterns, which could unintentionally erase Deaf speech characteristics that users consider meaningful markers of their identity. In this context, Deaf accent cloning was framed not as a corrective tool, but as a way for users to maintain authenticity, identity and signal community belonging.

\begin{figure}[ht]
    \centering
    \includegraphics[width=0.46\textwidth,alt={Interface concept for editing and controlling AI‑generated voice output. The layout includes segmented text blocks paired with tone tags, a sentence preview, and sliders for adjusting pitch, speed, and energy, illustrating how different attributes like emotions could be applied at the sentence level.}]{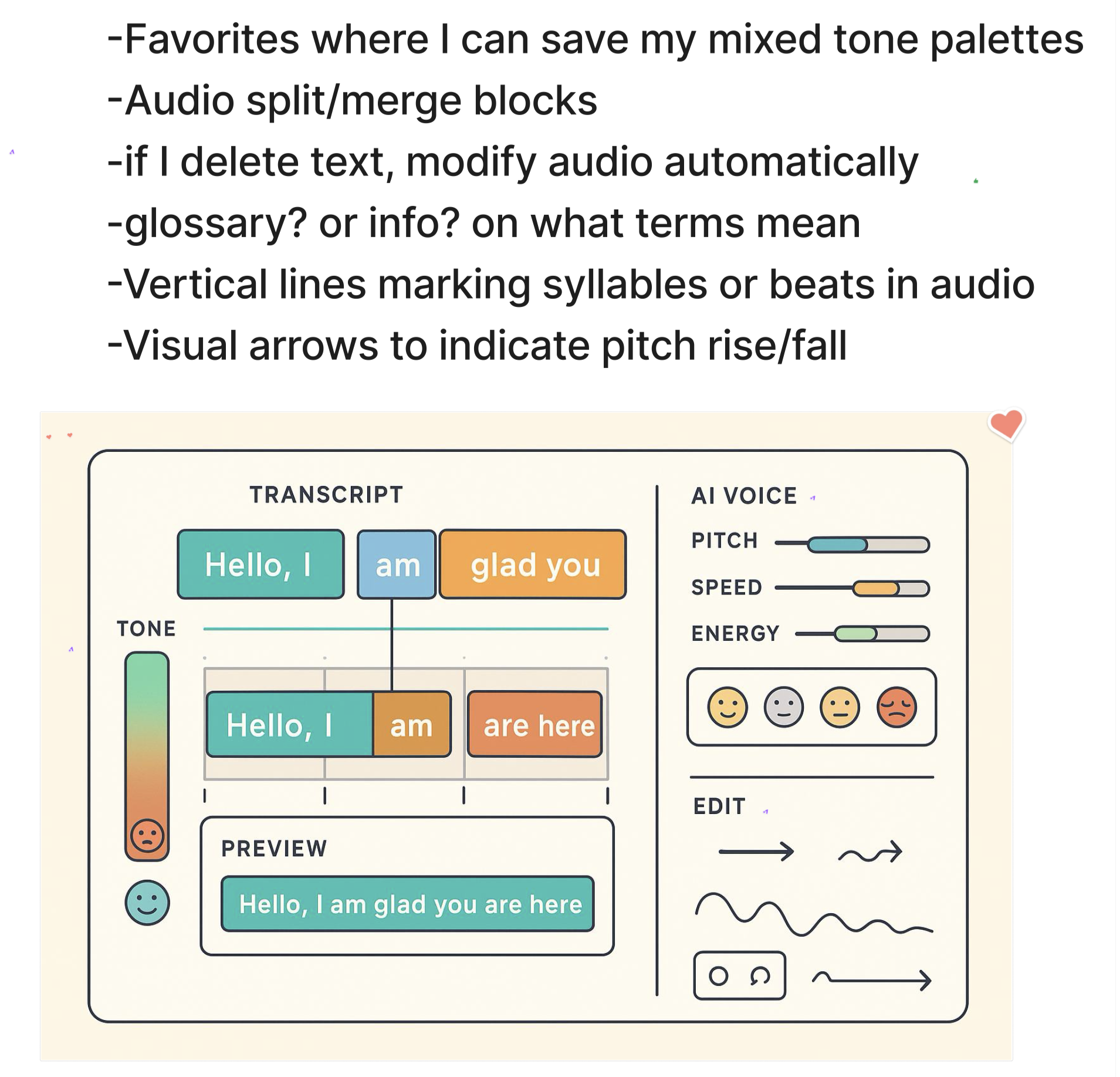}
    \caption{Based off the sketches from participants in the session, the research team made this creator design with ideas for adjustments and control and workflow.}
    \label{fig:workflow}
    \Description{Interface concept for editing and controlling AI‑generated voice output. The layout includes segmented text blocks paired with tone tags, a sentence preview, and sliders for adjusting pitch, speed, and energy, illustrating how different attributes like emotions could be applied at the sentence level.}
\end{figure}

The third co-design session broadened the scope by combining conceptual reflection with content creator perspectives. They contributed insights from content creators' lived experience online, explaining that most of their current followers are DHH because their content is primarily in sign language. While captions make their videos accessible to hearing viewers, standard captions fail to capture personality, tone, or emotional nuance. This limitation can make it harder to grow a broader audience, since hearing viewers may not feel the same connection to the content. These [DHH] creators suggested that TTS could help attract more hearing followers without erasing their identity expressed in signing by generating speech that carries both words and expressive intent. They emphasized the importance of considering workflow and proposed a design that embodies many of the control and editing themes also raised in the first two sessions (Figure~\ref{fig:workflow}). 

\begin{figure}[ht]
    \centering
    \includegraphics[width=0.4\textwidth,alt={Scenario‑wheel concept indicating how voice traits can be combined or contrasted. A central hub connects to labeled traits; celebratory, humorous, energetic, solemn, and productive, supporting reflection on how different emotional qualities interact and how they might influence pitch, pace, and volume across contexts.}]{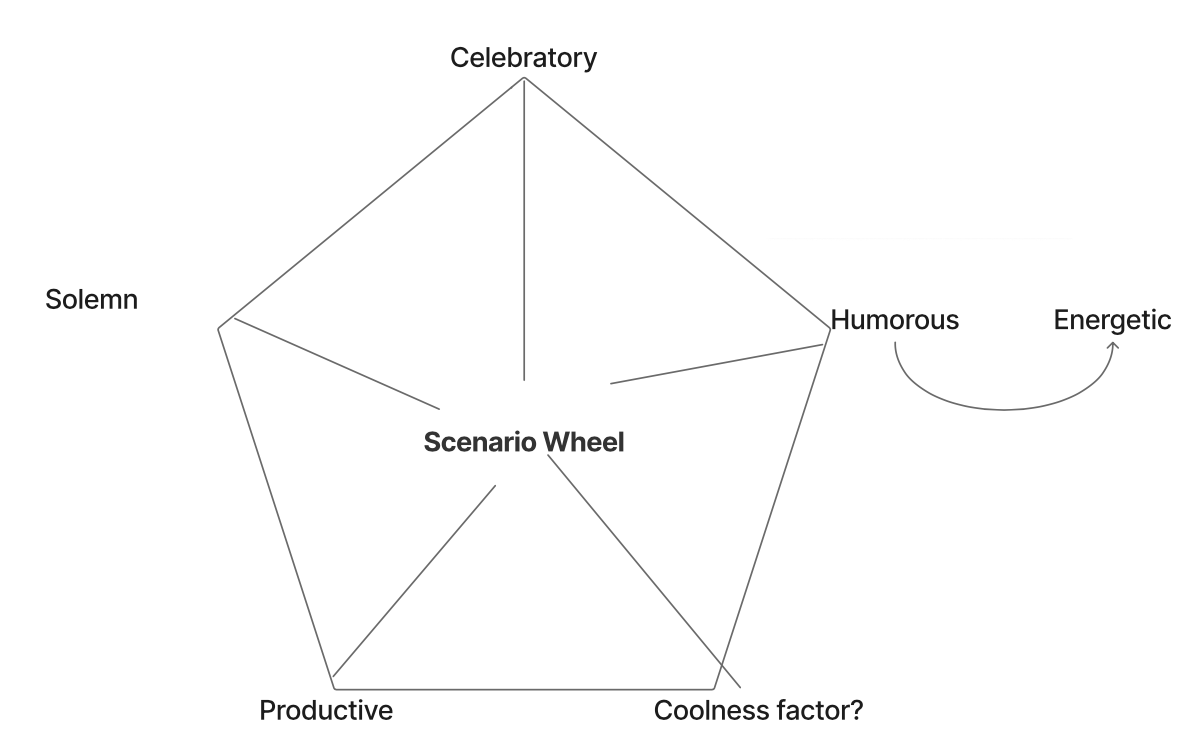}
    \caption{Scenario wheel for combing voice traits.}
    \label{fig:wheel}
    \Description{Scenario‑wheel concept indicating how voice traits can be combined or contrasted. A central hub connects to labeled traits; celebratory, humorous, energetic, solemn, and productive, supporting reflection on how different emotional qualities interact and how they might influence pitch, pace, and volume across contexts.}
\end{figure}

\begin{figure}[ht]
    \centering
    \includegraphics[width=0.4\textwidth,alt={Karaoke‑style caption concept where a red dot jumps across the text in a wave‑like path as the TTS voice speaks, providing a visual cue to indicate pitch, pace, and rhythm.}]{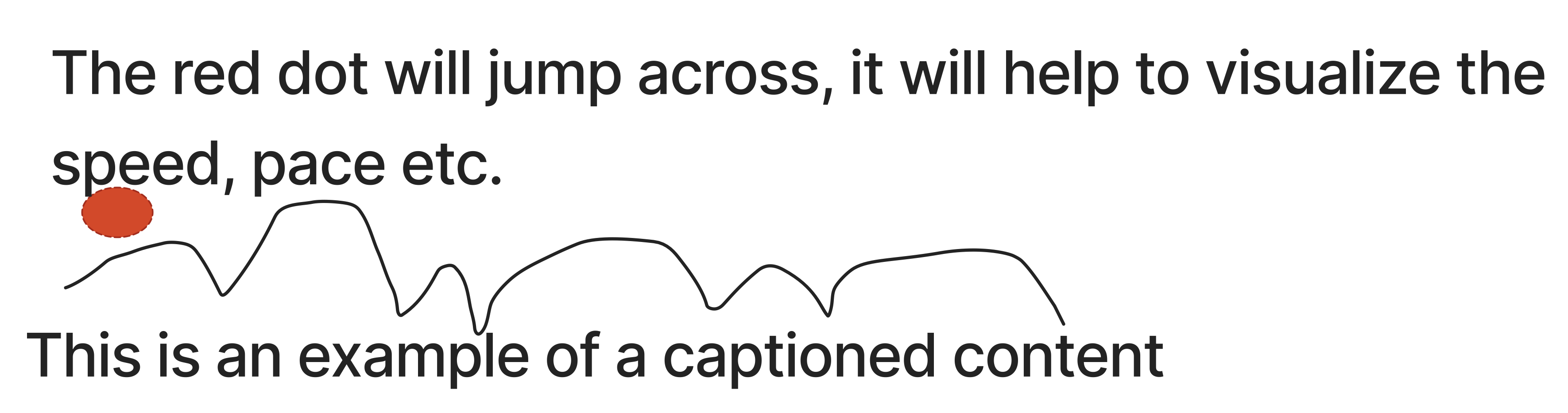}
    \caption{Karaoke captions with moving dot indicating rhythm and pitch.}
    \label{fig:karaoke}
    \Description{Karaoke‑style caption concept where a red dot jumps across the text in a wave‑like path as the TTS voice speaks, providing a visual cue to indicate pitch, pace, and rhythm.}
\end{figure}

Further, creators introduced scenario wheels (Figure~\ref{fig:wheel}) and description wheels to support consideration of how emotional traits combine or contradict, and how pitch, pace, and volume are perceived across contexts. Sign-to-voice translation was imagined as a speculative idea that could convert signing directly into appropriately delivered spoken output, bypassing TTS limitations. Creators further proposed karaoke captions (Figure~\ref{fig:karaoke}) and animated waveforms as verification tools, enabling users to visually track timing, pitch, and emphasis. An AI verification button was also proposed, mirroring the second session. Lastly, ensuring that personalized synthetic voices could be persistent across apps and social media platforms was seen as essential for building trust and maintaining authenticity.  

\subsection{Early Stage Design Evaluation Results}
\label{sec:oneonone}

\begin{figure}[ht]
    \centering
    \includegraphics[width=0.4\textwidth,alt={Screenshot of a Figma wireframe showing TTS voice preview card illustrating a highlighted word in a sentence tagged with an emoji to show excitement, with an emoji reference label at the top, and an AI voice summary command below the text.}]{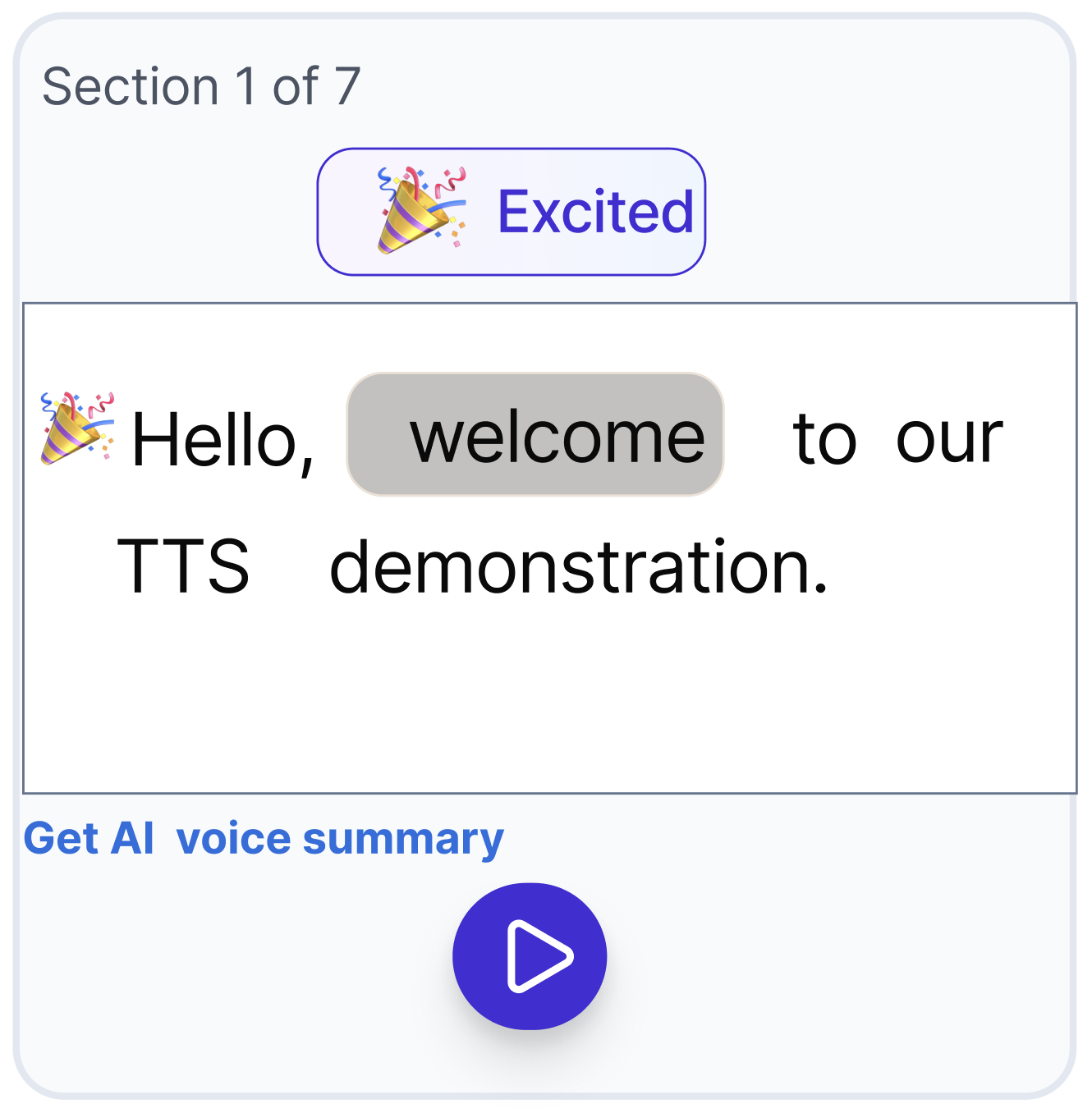}
    \caption{Emoji-based speech visualization design}
    \label{fig:emoji_figmadesign}
    \Description{Screenshot of a Figma wireframe showing TTS voice preview card illustrating a highlighted word in a sentence tagged with an emoji to show excitement, with an emoji reference label at the top, and an AI voice summary command below the text.}
\end{figure}

\begin{figure}[ht]
    \centering
    \includegraphics[width=0.4\textwidth,alt={Screenshot of a Figma wireframe showing TTS voice preview card with a red dot over a highlighted word in a sentence, and  an AI voice summary command. The layout illustrates how the red dot will move in sync with spoken texts.}]{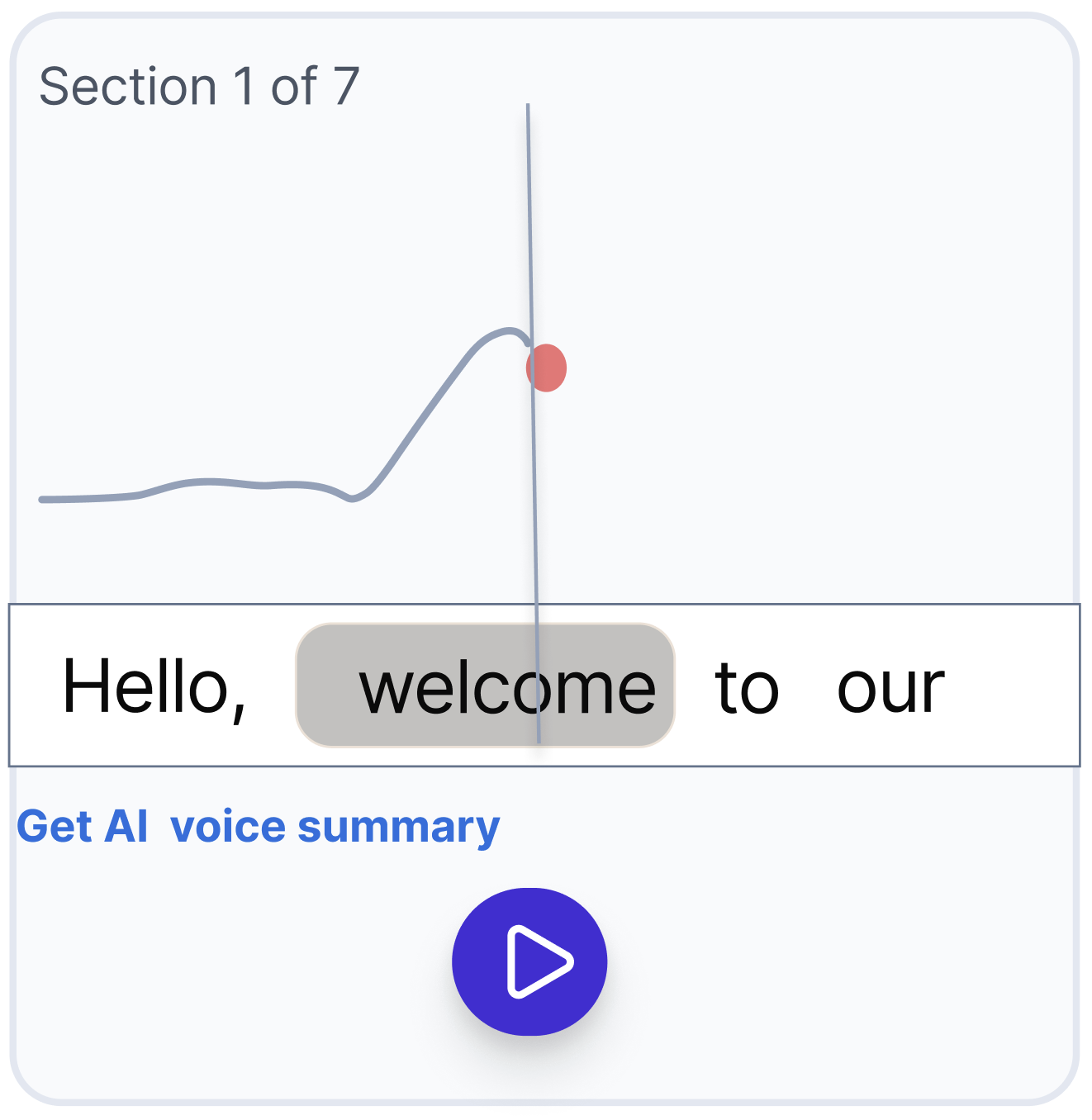}
    \caption{Karaoke-style speech visualization design}
    \label{fig:karaoke_figmadesign}
    \Description{Screenshot of a Figma wireframe showing TTS voice preview card with a red dot over a highlighted word in a sentence, and  an AI voice summary command. The layout illustrates how the red dot will move in sync with spoken texts.}
\end{figure}

\begin{figure}
    \centering
    \includegraphics[width=0.4\textwidth,alt={Screenshot of a Figma wireframe showing TTS voice preview card featuring a waveform made of vertical bars, with a text placed underneath it and an AI voice summary command positioned below the text. The bars indicate speech prosody during playback.}]{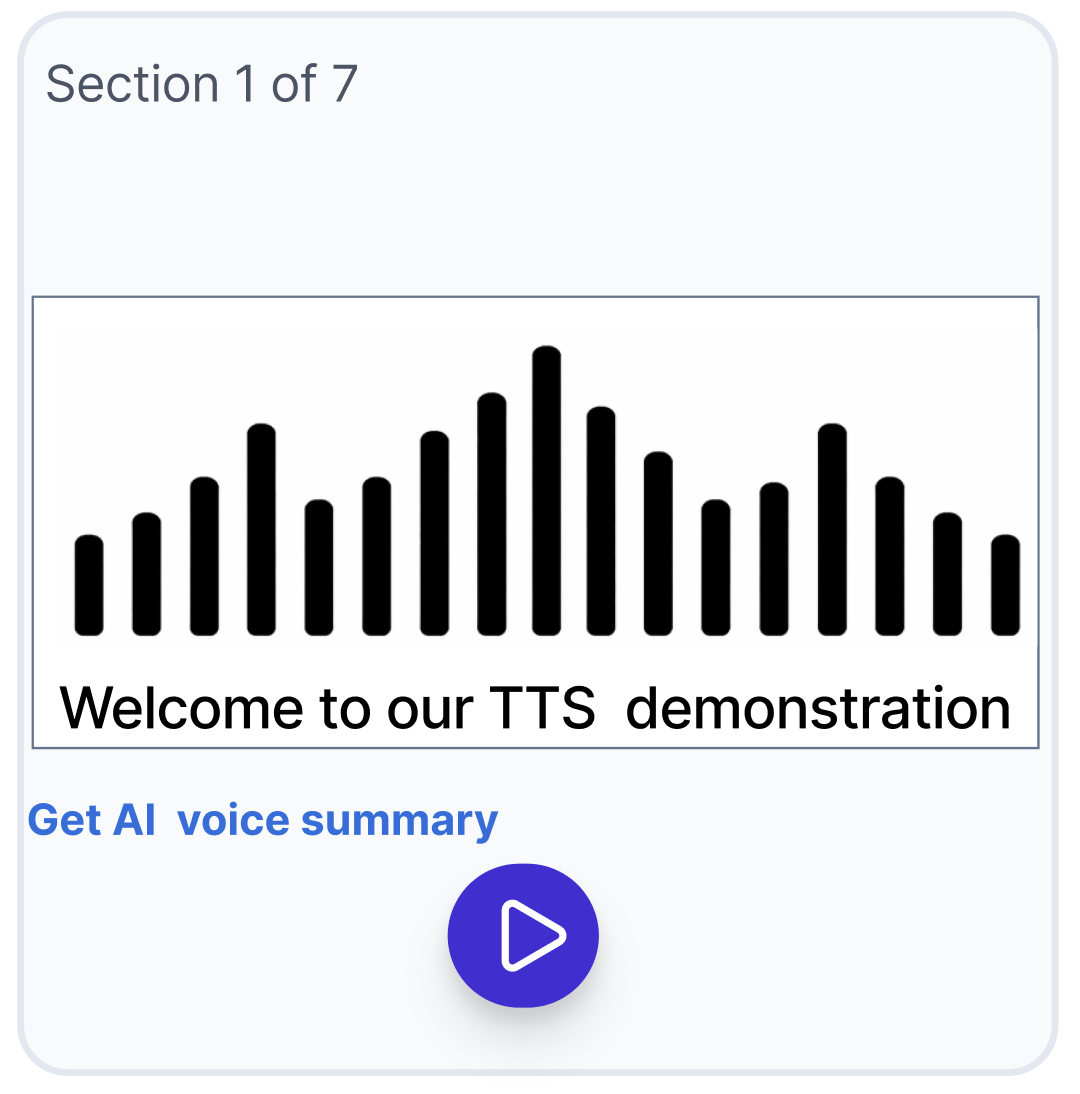}
    \caption{Waveform-style speech visualization design}
    \label{fig:waveform_figmadesign}
    \Description{Screenshot of a Figma wireframe showing TTS voice preview card featuring a waveform made of vertical bars, with a text placed underneath it and an AI voice summary command positioned below the text. The bars indicate speech prosody during playback.}
\end{figure}

The four one-on-one early-stage design‑evaluation sessions focused on validating how early visual representations of prosody, tone, and delivery supported DHH users’ ability to judge whether a TTS voice matched their intended communication style. Building on the 80/20 prioritization from the co‑design session, the evaluation centered on three core visualization concepts; emoji‑based, karaoke‑style, and waveform‑based representations; which reflected the most frequently recurring and high‑impact ideas for non‑auditory verification of TTS output. Results are organized according to the three evaluation phases described in the methods: (1) immediate interpretability, (2) understanding of tone and expressiveness through short demos, and (3) deeper comparison of tone, pitch, and emotion representation across designs. Across all phases, we assessed how effectively each concept supported the four design goals (cf. Section~\ref{sec:design-data-analysis}). The findings represent early insights that will be expanded through additional evaluation sessions with higher-fidelity prototypes.

\subsubsection{Immediate interpretability (five-second test)}
The Five‑Second Test captured participants’ rapid, intuitive impressions of each visualization. Across all four sessions, participants reported varying levels of initial clarity, with most expressing uncertainty about how the concepts functioned based on static wireframes alone.

\textbf{Emoji-based design} The emoji concept generated the most curiosity, with P4 describing it as “fascinating''. However, participants struggled to understand the functions or what the design was trying to communicate.

\textbf{Karaoke-style design} The karaoke concept produced the most uniformly uncertain responses. P2 acknowledged its visual appeal, but was unable to interpret the visualization. P3 reported feeling lost. P4 made a connection between the moving dot and vocal inflection by analogy to ASL body movement, but acknowledged this was speculative. Overall, the design's purpose was not discernible at first glance.

\textbf{Waveform-based design} Reactions to the waveform were more grounded, with P4 connecting it to pitch changes over time. However, P3 likened it to a fence or video player, and P2 found it overwhelming. While the waveform evoked some relevant mental models around audio, it did not yet clearly communicate its role as a TTS preview tool.

\subsubsection{Understanding of  Emotional Tone and Expressiveness Through Short Demos}
After viewing the short demo prototypes, participants developed a more nuanced understanding of each concept's strengths and limitations with respect to conveying tone and expressiveness. While initial reactions during the five-second test were largely driven by confusion about function, the demo phase shifted attention toward how well each visualization supported participants' ability to assess whether the TTS voice matched their intended communication style.

\textbf{Emoji-based demo} The emoji demo were generally received as the most immediately legible of the three concepts for conveying emotional tone. P4 described the experience positively, noting that the emojis changing in real time felt usable and gave a sense of control over tone. P1 noted that emojis were helpful for conveying emotions at the sentence level, but raised an important limitation: emojis lack universally agreed-upon meanings, which introduces ambiguity. They suggested that a dictionary or legend of emoji meanings would be necessary for the design to function reliably across users. P2 suggested that the system apply color coding to different tones and noted that sentence-level (rather than word-level) emotion assignments would reduce the user burden, as people less often shift emotional tone within a single sentence. Despite these concerns, the emoji concept was recognized across sessions as the most visually accessible approach to tone representation.

\textbf{Karaoke-style demo} Participant responses to the karaoke demo were the most negative across the evaluation. P2 described it as producing a complete disconnection from the material, rating the user experience as extremely poor and frustrating. They also emphasized that the design required too much reading, which disrupted the natural flow of interaction and felt unnatural. P1 noted that while the karaoke format convey emphasis, it felt more appropriate for entertainment or children's stories than for the kind of intentional, real-time tone verification that DHH users require. P3 found that the bouncing dot conveyed rhythm and pace but did not differentiate between emotional states, noting that excited and calm voices appeared visually equivalent in terms of dot behavior, rendering the design ineffective for conveying expressiveness. P4 echoed this, observing that the dot did not bounce with the intensity one would expect from a more emotional voice. Across participants, the karaoke design was seen as better suited for showing pace and rhythm than for conveying tone, expressiveness, or emotional nuance, which are central to the study's expressiveness and confidence design goals.

\textbf{Waveform-based demo} The waveform demos generated the most substantive and constructive feedback across all sessions. P1 identified the waveform as their personal favorite, noting that it was the only concept that broke down expressiveness at the word level rather than applying a blanket tone to a full sentence, a distinction they considered meaningful. P3 described the waveform with strong enthusiasm, noting that it offered a way to see pitch changes, rate of movement as a proxy for pace, and amplitude as a representation of volume. However, P3 also raised a critical concern: while the waveform conveyed volume well, it did not reliably distinguish between a loud and an excited voice, meaning that emotional interpretation still requires trust in the labels and settings accompanying the visualization rather than in the visual form itself. P4 appreciated that the waveform was very visually representative of tone and amplitude, and that its size and speed changes made prosodic variation legible. Across participants, the waveform concept was seen as the strongest of the three for conveying prosodic variation, though participants consistently identified the need for richer labeling and emotional annotation to reach its full potential.

Overall, the demo phase revealed that expressiveness and confidence in TTS output assessment are not supported by any single visualization alone. Participants across sessions gravitated toward the waveform for prosodic detail, the emoji for emotional clarity, and occasionally the karaoke for rhythmic pacing—suggesting that integration across concepts may be necessary to fully support DHH users' ability to verify that a generated voice matches their intent.

\subsubsection{Deeper Comparison of Tone, Pitch, and Emotion Representation Across Designs}
The side-by-side walkthrough elicited participants' most reflective and comparative assessments of the three design concepts. Presented with the full storyboard-style wireframes, participants were able to evaluate each concept in the context of a complete interaction flow, which surfaced both cross-cutting preferences and design-specific concerns related to the study's goals of clarity, expressiveness, confidence, and identity alignment.

While participants valued different elements of each design concept, no single design satisfied all four study goals. P4 found the waveform strongest for assessing voice characteristics but suggested integrating emojis for better emotional representation. P3 preferred a hybrid of the waveform and karaoke for judging pitch and pace. Conversely, P2 favored the emoji concept, noting that word-level highlighting helped them understand which word is being said, while also proposing sliders for adjusting emotional intensity. These diverging preferences underscore how communication needs directly influence which design elements are prioritized.

A primary concern across sessions was the difficulty of assessing emotional tone through visual cues alone. P3, who is late-deafened and relies entirely on visual design to calibrate the voice, noted that a high-amplitude waveform is ambiguous—potentially representing either excitement or yelling without accompanying labels. P1 also expressed interest in real-time facial expression recognition during typing, though they cautioned that such a feature must account for cultural nuance, noting that expressions can be interpreted differently across different countries or cultures.

Confidence in these designs remains tied to the tension between control and efficiency. P3 noted a lack of confidence without a way to independently verify the voice’s accuracy. Meanwhile, P2 expressed a pragmatic preference for a ``close enough'' voice that required only one or two clicks, emphasizing speed over fine-grained control.

\section{Discussion}

Our findings extend several threads identified in the related work. Prior research on emotional TTS has shown that systems can capture broad affective categories but remain constrained by simplified taxonomies and auditory-only evaluation \cite{aoki2022clear,gessinger2022cross,shaikh2010improving,zhou2022speech}, a limitation that our participants' emphasis on non-auditory verification relates to directly. Speaker-adaptive TTS approaches have similarly demonstrated the ability to preserve aspects of vocal identity \cite{biadsy2019parrotron}, \cite{jia2018transfer}; our findings extend this work by reframing identity preservation as a design value defined by DHH users themselves, rather than a purely technical benchmark. Our co-design methodology also builds on established practices for participatory work with DHH communities \cite{may2024co,mcdonnell2023easier,seita2022remotely}, extending those practices into the previously unaddressed domain of TTS. Across all phases of this research, participants revealed how DHH users envision TTS not merely as a communication aid, but as a way to express identity, convey emotions, and manage interactions. Their insights collectively exposed the structural gaps in current TTS systems and clarify that Deaf-centric design must achieve non-auditory verification methods, expressive customizations, and authentic representation. 

\subsection{Non-auditory Verification as a Fundamental Design Requirement}
A consistent motif across all phases of this work is the problem of verification: how can DHH users assess whether TTS output sounds appropriate, natural, or emotionally aligned with their intent, when they cannot reliably access the audio? This challenge is not merely a usability inconvenience but a design challenge. As our focus group participants illustrated, one participant had unknowingly been producing poor-quality speech without realizing it, with potential social consequences in hearing interactions. Current TTS evaluation frameworks offer no solution here: as noted in the related work, emotional TTS research relies exclusively on auditory metrics~\cite{aoki2022clear,gessinger2022cross,shaikh2010improving,zhou2022speech}, which are inaccessible to many DHH users by definition.

Our co-design sessions produced a range of proposed solutions: waveform overlays, emoji cues, karaoke-style speech visualization, AI-generated tone summaries, and crowdsourced feedback. The design evaluation allowed us to assess these various approaches. No single visualization proved sufficient. The waveform conveyed prosodic detail at the word level but could not reliably distinguish emotional states such as excitement from loudness without additional labeling. Emojis were the most immediately legible for emotional tone but introduced ambiguity, with participants calling for a definitional legend. The karaoke format was better suited for rhythm and pace than for emotional nuance. These findings suggest that effective non-auditory verification may require integration across visualization types. This is a contribution that extends beyond the DHH context: the lack of non-auditory evaluation mechanisms is a structural gap in TTS research more broadly~\cite{aoki2022clear,gessinger2022cross}, and Deaf-centric design foregrounds it in a way that hearing-centric design has not provided.

\subsection{Identity, Representation and the Limits of Generic Voice Designs}
A second major finding is that DHH users require TTS systems to go well beyond functional speech generation by supporting authentic self-representation. Focus group participants wanted voices that reflected intersectional identities, including gender expression, community belonging, and situational context, echoing prior work on the social and communicative weight of voice characteristics~\cite{lee2021speech,leongomez2021voice}. The co-design sessions extended these into possible designs; voice profile builders with cultural and demographic markers, preservation of Deaf accent (through voice cloning), and layered emotion tagging were proposed as mechanisms to ensure that synthetic voices do not erase individuality. This responds directly to a gap in the speaker-adaptive TTS literature, which has demonstrated the ability to retain aspects of vocal identity~\cite{biadsy2019parrotron,jia2018transfer}, but has not been tested for DHH-accented speech. Our participants' insistence that Deaf-accented speech be preserved, rather than corrected suggests that successful identity retention in this context should be defined by fidelity to the users own voice, rather than by conformity to normative speech pattern.

Our work suggests that accent preservation is not merely a technical challenge but a design value; participants in the second co-design session explicitly identified the preservation of Deaf-accented speech as a critical requirement, not a problem to be corrected. This distinction between challenge and value matters and should inform how DHH voice cloning is framed in future research. The design evaluation further reinforced that identity alignment is not well-served by any current visualization concept, since emotional interpretation remains tied to trust in labels and system settings rather than the visual form itself, pointing to the need for richer annotated and user-configurable labeling systems.

\subsection{Usability, Workflows and the Diversity of DHH Use Cases}
The focus groups also revealed that participants’ primary envisioned use case was interactive, real-time communication, situations where TTS latency and social perception are significant barriers, consistent with prior findings on communication delay in DHH technology contexts~\cite{napier2011difficult,rui2022online,vogler2013mixed}. This initial narrow frame was productive: it surfaced genuine adoption barriers that a researcher-driven frame might have missed.

However, introducing personas in the second co-design session and recruiting content creators for the third broadened the design space considerably, consistent with recommendations from prior DHH co-design work~\cite{may2024co,mcdonnell2023easier,sales2020participatory}. Content creators brought a distinct set of priorities, workflow integration, platform continuity, and expressive tools for reaching hearing audiences, that align with recent findings on DHH streamers and TikTokers~\cite{cao2023sparkling,cao2024voices}. Their insight that captions alone fail to convey personality, tone, and emotional nuance to hearing audiences, and that TTS could bridge this gap without erasing signing identity, represents an important contribution to understanding TTS as more than an accessibility accommodation. It positions TTS as a creative and professional tool with implications for how DHH content creators can participate in mainstream digital spaces.

The design evaluation added a further dimension, revealing a tension between the desire for fine-grained control and the practical need for speed: some participants wanted word-level precision while others prioritized a ``close enough'' result with minimal interaction. This argues for layered interfaces that offer both low-complexity entry points, presets, sliders, emotional wheels, and deeper editing capabilities to meet diverse DHH needs.

\subsection{Ethics, Trust, and Voice Cloning}
Participants raised concerns about fairness, accountability, transparency, and ethics (FATE) in relation to voice cloning and AI-generated verification. These concerns are grounded: the ability to replicate a voice from an audio sample raises real risks of misrepresentation, non-consensual use, and reinforcement of stereotypes. These are concerns that participants connected directly to their own vulnerability as DHH people whose voice identity can influence how seriously they are taken.

Privacy concerns were also raised around how cloned voices would be saved, accessed and used. Safeguards methods were proposed including explicit consent mechanisms, transparency about synthetic voice use, and accountability structures for misuse. While FATE considerations have been examined for sign language AI datasets \cite{bragg2021fate}, no comparable framework exists yet for Deaf-centric voice technologies. This work surfaces the need for one and provides an initial set of user-generated design requirements around consent and transparency that future work can build on.

\section{Technical Requirements for Implementing the Designs}
\label{sec:tech-reqs}

The outcome of our participatory design process places a number of technical demands on both TTS systems specifically and AI more generally. These must be met before the designs in this paper can become a reality. We organized these requirements into participant-driven needs and researcher-interpreted insights.

\subsection{Participant-Driven Needs}
Participants expressed needs related to adjusting emotions on a continuum, shifting tone between formal to informal speech, using context-dependent presets such as for Doctor visits, and applying per-word emotion tagging. they also emphasized strict user privacy, Deaf-accented voice options (both removing and retaining voice characteristics), and non-auditory verification of generated speech.

\subsection{Researcher-Interpreted Insights}
In the following we list topics that have been identified in consultation with a TTS expert. 

\subsubsection{Emotion Customization}
Adjusting emotions on a continuum will require Valence-Arousal-Dominance (VAD) modeling~\cite{guoping2025vad}. There are challenges with mapping VAD to an intuitive UI. Adjusting tone between formal to informal speech on a continuum also could be challenging, as technology typically would be trained on discrete tone states; likewise, additional work is required to blend both tone and emotions. Context-dependent presets, such as job interviews, will need to be mapped to domain-specific TTS/voice data. Per-word, as opposed to per-sentence, emotion tagging requires annotated training data that has emotions annotated separately for each word, rather than entire sentences or passages. It is unclear whether such data currently exists. Strict user privacy implies running TTS locally on a device, rather than the cloud. However, generative AI-based TTS models are currently too large to fit on many consumer devices. 

\subsubsection{Voice Cloning}
Deaf-accented voice cloning that removes the accent characteristics is still uncharted territory, but likely can be done via adapting standard voice cloning techniques, such as zero-shot cloning. However, deaf-accented voice cloning that retains specific identifying deaf-accented characteristics is a huge challenge, because typical cloning is based on datasets of normative speech accents and characteristics. Variable DHH accents would need to be included in such datasets first. 

\subsubsection{AI Prerequisites}
Facial expression recognition support will be essential to enable emotion tagging based on the users' face while typing. If emotion and tone is to be conveyed through signing style, on top of facial expressions, significant advances in sign language recognition are needed. Conversely, for conveying emotions and tone in ASL to a user, significant research into sign language avatars will be necessary. At present there are none that can sufficiently convey the detailed nuances of emotions on top of grammatical markings in ASL~\cite{mcdonald2025emotion}. There are also gaps in translating emotion-tagged speech or writing into functionally equivalent ASL expressions, even if avatars were capable of rendering them.

\subsubsection{Supporting AI-based Verification}
Supporting AI-based verification of generated speech requires accurate recognition of emotion, tone and delivery in ASR, which is still an area of active research~\cite{he2024speechemotion}. Even if ASR were capable of recognizing these attributes reliably, there are open questions as to how these could be translated into written English or ASL in a way that is readily understandable. Picking appropriate voice characteristics based on context, desired tone and delivery, or personas, on the other hand, is close to being a reality with existing large language models. Commercial implementations exist, such as ElevenLabs's. However, fine-tuning such models to take the unique cultural characteristics of DHH people into account, is likely to be beneficial.

\section{Overall Implications and contributions}
This work makes the following contributions to the accessible HCI and TTS research communities:
\begin{enumerate}
    \item It establishes participatory design with DHH users as a viable and productive methodology for TTS development, extending prior co-design work in the DHH space \cite{may2024co,mcdonnell2023easier,sales2020participatory,seita2022remotely} into a domain that has not previously been addressed this way.
    \item It surfaces non-auditory verification as a foundational, and currently unmet, design requirement for Deaf-centric TTS, and provides an empirical evaluation of three candidates for visualization approaches, offering a basis for future prototyping. A key theme from the design evaluation is that visualization is challenging, and our exploration has not yet found a good approach. Tentatively, a combination of several approaches may be effective.
    \item It reframes Deaf accent preservation as a design value rather than a technical problem, with implications for how speaker-adaptive TTS systems \cite{biadsy2019parrotron,jia2018transfer} should be developed and evaluated for DHH users.
    \item It expands the range of real world applications for DHH TTS beyond interactive communication to include content creation and professional media production, with implications for how tools like those described by Cao et al. \cite{cao2023sparkling,cao2024voices} might be designed.
    \item It identifies voice cloning ethics as an open research and design problem in the DHH space, establishing a set of user-grounded requirements that complement theoretical FATE frameworks such as those developed for sign language AI \cite{bragg2021fate}.
    \item It identifies areas where further foundational research and development is needed, especially in the areas of identity-preserving speech synthesis and sign language technologies. 
\end{enumerate}

\section{Limitations}
\label{sec:limitations}
With the four different example personas during the co-design session, we considered a diversity of communication modalities among potential TTS users. However, we did not include any DeafBlind or DeafDisabled personas. This is something that was pointed out by a co-design participant and will be critical to consider as we continue developing our project. We also had limited geographical reach. The majority of co-design participants were local to Washington DC, the area where the study was conducted. Only the focus groups and one co-design session were conducted virtually with participants from across the United States. An additional limitation is that all participants were based in the U.S., and thus our results reflect a Western-centric viewpoint.

A further limitation relates to the early-stage design evaluation. The short demo prototypes were not interactive. Participants viewed brief demonstrations of the output screen for each visualization concept but could not manipulate controls, generate speech, or experience the full setup‑to‑output workflow of a functional TTS system. Their feedback therefore reflects interpretation of the concepts rather than hands‑on use. The evaluation also included only four participants, which limits our ability to draw conclusions about a preferred design direction.

Another limitation is that significantly more work on managing the risks and promise of AI in conjunction with TTS is needed. Uncritical applications of AI could result in perpetuation of stereotypes and stigma around DHH people as well as intersectional identities, particularly with respect to their use of speech. On the flip side, unrealistic expectations of what AI is capable of now and and will be capable of in the future, also carries risks. For example, participants suggested the use of facial expression recognition for integrating tone into TTS voice output. Similarly, they suggested using AI to apply tone and verify that the tonal output matches the intent of the user. Current technological capabilities make both of these difficult to implement (see also Section~\ref{sec:tech-reqs}), and even if future implementations become practical, biases and insufficient diversity in AI training data may preclude their application in DHH contexts.

\section{Future Work}
Future work will further refine the design space, with additional design evaluations and co-design sessions, as well as implementing the design concepts in concrete prototypes that allow hands-on testing. Collective evaluation of all the design concepts that we have presented will be another important future direction. Such an evaluation would allow participants to compare the designs side‑by‑side, discuss trade‑offs together, and help determine whether one design, or a combined approach best meets the goals of Deaf‑centric TTS.

Some areas also require dedicated research and development to make the participants’ co-design ideas technically feasible. Emotion detection in speech remains an active research frontier, with challenges in reliably capturing subtle effective cues across diverse voices, and context. Constructing synthetic voices with AI and large language models is another critical area, as systems must learn to balance naturalness, personalization, situational context, and safety. Finally, sign language recognition to drive TTS is still a long way off. This requires advances in machine learning and culturally sensitive translation pipelines. 

\section{Conclusion}
This work establishes a foundational design framework for Deaf‑\linebreak centric TTS systems by grounding identity, authenticity, and non‑\linebreak auditory verification in practical insights from focus groups, co‑design, and design evaluation with DHH participants and creators. Through these methods, we have surfaced a layered set of requirements: simplicity through low‑complexity entry points, identity‑driven customization that preserves rather than erases Deaf identity, and verification mechanisms that allow users to visually confirm alignment with communicative intent. The empirical evaluation of visualization concepts grounds these priorities in concrete design trade‑offs. 

Together, these findings lay the groundwork for a new area of DHH technology research and establish participatory design as a productive path toward TTS systems that are genuinely built for and with the communities they are meant to serve. This is uncharted territory, and the first work of its kind. We do not claim to offer a complete solution; however, this work should serve as a set of guiding principles for the creation of future TTS technologies that are inclusive of DHH individuals, as well as a reminder to always design with those we are designing for.

\section*{AI Use Statement}
We used AI-assisted tools during prototype preparation. Figma Make was used to automatically translate our designed wireframes into basic interaction flows for the Emoji-style prototype. OpenAI Codex was used to translate the Karaoke and Waveform wireframes into lightweight web-based short demo prototypes, including simulating typing text and previewing synthesized speech. These tools supported rapid prototyping only, all study design decisions, research materials, analyses, and interpretations were developed by the authors. No generative AI tools were used to analyze data or write any part of this paper.

\begin{acks}
The contents of this paper were developed under a grant from the National Institute on Disability, Independent Living, and Rehabilitation Research (NIDILRR grant number 90REGE0027). NIDILRR is a Center within the Administration for Community Living (ACL), Department of Health and Human Services (HHS). The contents of this paper do not necessarily represent the policy of NIDILRR, ACL, HHS, and you should not assume endorsement by the Federal Government. Additional support has been provided by the National Science Foundation under Award No. 2440601. Matthew Seita assisted with the planning and execution of the co-design sessions. Abdifatah Mohamed and Lauren Medina supported the focus groups. Alex Pérez at AppTek provided his expertise on the capabilities of TTS systems, which helped frame the section on technical requirements.
\end{acks}

\bibliographystyle{ACM-Reference-Format}
\bibliography{references}

\appendix

\end{document}